\documentclass[aps,prl,showpacs,twocolumn]{revtex4}
\usepackage{amsmath}
\usepackage{graphicx}
\def\){\right)}
\def\({\left(}
\def\]{\right]}
\def\[{\left[}

\newcommand{\be}{\begin{equation}}
\newcommand{\ee}{\end{equation}}

\newcommand{\lsim}{\, \, \raisebox{-0.8ex}{$\stackrel{\textstyle <}{\sim}$ }}

\newcommand{\roughly}[1]%
{\mathrel{\raise.4ex\hbox{$#1$\kern-.75em\lower1ex\hbox{$\sim$}}}}

\newcommand\beq{\begin{eqnarray}}\newcommand\eeq{\end{eqnarray}}

\def\Dsl{\,\raise.15ex \hbox{/}\mkern-12.8mu D}

\def\fm3{fm$^{-3}$}

\begin{document}
%\begin{frontmatter}
%
%\preprint{\vbox{\hbox{LA-U R- ?????}}}

\title{The Strange Star Surface: A Crust with Nuggets}

\author{Prashanth Jaikumar$^{\S}$, Sanjay Reddy and Andrew W. Steiner}

\affiliation{$^{\S}$Physics Division, Argonne National Laboratory, Argonne, IL 60439 USA\\
Theoretical Division, Los Alamos National Laboratory, Los
Alamos, NM 87545 
}
\begin{abstract}
We reexamine the surface composition of strange stars. Strange quark stars are hypothetical compact stars which could exist if strange quark matter was absolutely stable. It is widely accepted that they are characterized by an enormous density gradient ($~10^{26}$ g/cm$^4$) and large electric fields at surface.  By investigating the possibility of realizing a heterogeneous crust, comprised of nuggets of strange quark matter embedded in an uniform electron background, we find that the strange star surface has a much reduced density gradient and negligible electric field. We comment on how our findings will impact various proposed observable signatures for strange stars.  
\end{abstract}
\maketitle

%\section{Introduction}
%==============================================
The conjecture that matter containing strange quarks could be absolutely stable is several decades old~\cite{Itoh:1970jb,Bodmer:1971we,Witten:1984rs,Farhi:1984qu}. In the intervening years numerous authors have investigated how such matter would manifest in nature (for recent reviews see Refs.~\cite{Madsen:1998uh,Horvath:2003sb,Weber:2004kj}). Possibilities include tiny (dimensions of a few Fermi) quark lumps called strange nuggets and large (dimensions ~km) compact stars made entirely of strange quark matter~\cite{Baym:1976yu,Haensel:1986qb,Alcock:1986hz}. Strange stars, as discussed to date, have bulk homogeneous quark matter containing up, down and strange quarks extending all the way to the surface, which is uniquely qualified by (i) a steep density drop $\Delta \rho \sim  ~10^{15}$ g/cm$^3$ over a distance of several Fermi; and (ii) large electric fields due to less rapid variation of the electron density~\cite{Alcock:1986hz}. In this letter, we reexamine the surface region of these strange stars and find that, contrary to conventional wisdom, a heterogeneous (solid) crust made of strange nuggets and electrons is likely, leading to a much reduced density gradient and negligible electric fields in the surface region. This is very different from the conventional picture, suggested by Alcock et al~\cite{Alcock:1986hz}, of a tiny nuclear crust suspended a few hundred Fermis above the quark star, supported by large electric fields near the surface. Our proposal shares some apparent features of the quark-alpha crust scenario, based on the stability of tightly bound, low-baryon and absolutely stable strange quark states called quark-alphas~\cite{Benvenuto:1990,Michel:1988xp}, but it is otherwise fundamentally different.  

In the vicinity of the strange star surface hydrostatic equilibrium requires the pressure to become vanishingly small. At the surface the pressure is identically zero. The pressure of stable quark matter vanishes at a finite and large quark density $n\simeq 1$ quark/fm$^3$, corresponding to a quark chemical potential of $\mu \simeq 300 $ MeV.  Since the strange quark mass, $m_s$, is large compared to the up and down quark masses homogeneous quark matter needs electrons to ensure charge neutrality. In normal (non-superconducting) quark matter the electron chemical potential needed to ensure neutrality is $\mu_e \simeq m_s^2/4\mu$.  Consequently, when the total pressure ($P_{\rm quark} + P_{\rm electron}$) is close to zero, the pressure due to quarks is negative. We show that in this regime a heterogeneous mixed phase with nuggets and electrons may be favored if surface and Coulomb costs are small.  This mixed phase  is qualitatively similar to the mixed phase of nuclei and electrons in the crust of normal neutron stars and shares several features with the mixed phase of quark drops and nuclear matter in hybrid stars \cite{Glendenning:1992vb}. As in these other cases, the size of  nuggets in the mixed phase will be determined by minimizing the surface, Coulomb and other finite size contributions to the energy.  At low temperature, the mixed phase will be a solid. Using typical quark model parameters, we find that strange stars will have a relatively large crust with radial extent $\Delta R \simeq 50$ m for a  star with mass $M \simeq 1.4 M_{\odot}$ ($M_{\odot}\simeq 2\times 10^{33}$ g is the mass of the sun) and radius $R=10$ km.  The electron density decreases to zero over this length scale. 

To prove that a heterogeneous phase is favored when surface and Coulomb energies are negligible, we adopt a model independent approach which is valid when $\mu_e \ll \mu$. In this case, the quark pressure may be expanded in powers of $\mu_e$ and to second order in $\mu_e$, it is given by
\begin{equation}
P_{\rm q}(\mu,\mu_e) = P_0(\mu) - n_{Q}(\mu) ~\mu_e + \frac{1}{2} ~\chi_{Q}(\mu)~\mu_e^2 \,,
\label{eqn:Pressure}
\end{equation} 
\noindent where $n_Q(\mu)=- \partial P/\partial \mu_e$ is the positive charge density, $\chi_Q(\mu)=\partial^2 P/\partial \mu_e^2$ is the charge susceptibility and $P_0$ is the pressure of the electron-free quark phase. They depend on $\mu$, the strange quark mass $m_s$ and strong interactions. To perform a model-independent analysis we will treat $P_0$, $n_Q$, and $\chi_Q$ as ($\mu$-dependent) parameters. To appreciate their typical magnitude, we note that in the Bag model description $n_Q = m_s^2 \mu/2\pi^2 $, $\chi_Q = 2 \mu^2/\pi^2$ and 
 $P_0 = 3 (\mu^4-m_s^2\mu^2)/4 \pi^2 - B$, where $B$ is the bag constant. To investigate the regime where the electron contribution to the pressure is relevant, we should keep terms up to fourth order in $\mu_e$.  However, this greatly complicates the analytic treatment and does not provide much new insight. For the time being, assuming the $\mu_e^3$ and $\mu_e^4$ terms in the quark pressure to be numerically small compared to the electron pressure (which we include explicitly), we work to second order in $\mu_e$ and return to a more complete treatment of the problem later. 
 
 A heterogeneous state of positively charged quark matter coexisting with negatively charged electron gas is possible if $P_q=0$ and $\partial P_q/\partial \mu_e \le 0$.  Since electrons reside both inside and outside quark matter, Gibbs phase equilibrium requires $P_q=0$. We require $\partial P_q/\partial \mu_e \le 0$ to ensure that quark matter is positively charged - which is necessary to satisfy global charge neutrality.  
At fixed $\mu$, from Eq.~\ref{eqn:Pressure} we see that $P_q$ is zero and quark matter is positively charged when $\mu_e$ takes on the value 
\begin{equation}
\label{eqn:muecritical}
\tilde{\mu}_e=\frac{n_Q}{\chi_Q}~(1-\sqrt{1-\xi})\quad
{\rm where}~\xi=\frac{2P_0\chi_Q}{n_Q^2}\,.
\end{equation}
Hence a mixed phase is possible when $0<\xi < 1$. In this regime, the mixed phase has lower free energy (larger pressure)  than homogeneous matter. Relaxing the condition of local charge neutrality allows us to reduce the strangeness fraction in quark matter and thereby lower its free energy. $\xi=1$ characterizes the critical point where this becomes possible. The pressure at the critical point $P_c=(\mu_{e}^c)^4/12\pi^2$, where $\mu_e^c=n_Q/\chi_Q$ is the electron chemical potential there. In this phase, electrons contribute to the pressure while quarks contribute to the energy density - much like the mixed phase with electrons and nuclei in the crust of a conventional neutron star. 

Although $P_0$, $n_Q$ and $\chi_Q$ all depend on $\mu$, and consequently change across the mixed phase we find that only the variation in $P_0$ is relevant because $P_0$ varies rapidly with $\mu$ inside the mixed phase. For example, in the bag model, $\mu$ changes by less than a percent across the  
interval  $0\le \xi \le1$ so that to a good approximation, we can treat 
$n_Q$ and $\chi_Q$ as constants throughout the mixed phase. Further, since $\mu$ is nearly constant across the mixed phase the variation of the energy density inside nuggets is negligible. In what follows,  $\epsilon_0$ denotes the energy density inside nuggets.  

To characterize the mixed phase we need to determine how the electron chemical potential and the volume fraction of the quark phase change with $\xi$. We have already obtained Eq.~\ref{eqn:muecritical} which determines how $\mu_e$ changes with $\xi$. The volume fraction of the quark phase, denoted by $x$, is determined by the condition of global charge neutrality $Q(\tilde{\mu}_e)~x = n_e( \tilde{\mu}_e)$ where $Q(\tilde{\mu}_e)=-\left( \partial P_q/\partial \mu_e \right)_{\mu_e=\tilde{\mu}_e}$ is the quark charge density in the mixed phase. We find
\begin{equation}
x=\frac{\tilde{\mu}_e^3}{3\pi^2~n_Q}~\left(1-\frac{\chi_Q\tilde{\mu}_e}{n_Q}\right)^{-1}\,.
\label{eqn:volfrac}
\end{equation}

The mixed phase will be penalized by Coulomb, surface, and other finite size contributions to the energy. Its stability at fixed pressure is guaranteed if its Gibbs free energy (per quark) is lower  than the homogeneous phase.  The Gibbs energy per quark $g=(E+PV)/N$,  where $E$ is the energy, $P$ is the pressure and $N$ is the number of quarks in volume V. In the homogeneous phase, $g_{\rm H}=\mu_{\rm H}$ where $\mu_{\rm H}$ is the quark chemical potential. Similarly, $g_{\rm M}=\mu_{M}$ in the mixed phase if finite size contributions are neglected. We now calculate the Gibbs free energy gain $\Delta g=\mu_{\rm H}-\mu_{\rm M}$ in the mixed phase. 

Using the local charge neutrality condition $\mu_e=n_Q/\chi_Q$ and Eq.~\ref{eqn:Pressure}, we find the pressure of the homogeneous phase
\begin{equation} 
P_{\rm H}(\mu) = P_0(\mu) - \frac{1}{2} ~\frac{n_Q^2}{\chi_Q}\,.
\end{equation} 
The Gibbs energy $\mu_{\rm H}$ at fixed total pressure $P$  is then determined by the equation $P_0(\mu_{\rm H})= P + n_Q^2/2\chi_Q$. Since we expect $\Delta g \ll \mu$, a Taylor series expansion of the form  
\begin{equation} 
P_0(\mu_{\rm H})=P_0(\mu_{\rm M}) 
+n~\Delta g  + {\cal O}( \Delta g^2\mu^2)\,,
\end{equation} 
where  $n=\left(\partial P_0/\partial \mu\right)_{\mu=\mu_{\rm M}}$ is justified. When $\mu_e \ll \mu$, $n$ is the quark number density inside nuggets. Using Eq.~\ref{eqn:muecritical} and $\tilde{\mu}_e=(12\pi^2~P)^{1/4}$, the gain in Gibbs  energy per quark is 
\begin{equation} 
\Delta g= \frac{n_Q^2}{2 \chi_Q n}~\left( 1-\frac{2 \chi _Q \tilde{\mu }_e}{n_Q}+\frac{\chi _Q^2 \tilde{\mu
   }_e^2}{n_Q^2}\right)\,.
% +\frac{\chi _Q \tilde{\mu}_e^4}{6 \pi ^2 n_Q^2}
\label{eqn:Gain}
\end{equation}
In the bag model when $P=0$ , $\Delta g=  m_s^4/16 \pi^2n$. For $m_s =150$ MeV and $n=1/$fm$^3$, $\Delta g \simeq 0.4 $ MeV per quark.  The surface and Coulomb energy cost in the mixed phase has been studied in the context of the nuclear mixed phase~\cite{Ravenhall:1983uh}. Using these results, which are valid when corrections due to Debye screening and curvature energy are negligible, we find that the Coulomb and surface energy cost per quark
\begin{equation} 
\epsilon_{s+C}=\frac{6\pi}{n~(16 \pi^2)^{1/3}}~\left[(e^2~\sigma ~{\rm d} ~n_Q)^2~f_d(x)\right]^{1/3}\,,
\label{eqn:Loss} 
\end{equation} 
where $\sigma$ is the surface tension, d is the dimensionality (d=3 for spheres, d=2 for rods, and d=1 for slabs), and the function $f_d(x)$ depends on the dimensionality and the volume fraction $x$ of the rarer phase. Explicit forms for $f_d(x)$  may be found in Ref.~\cite{Ravenhall:1983uh}. From Eq.~\ref{eqn:Gain} and Eq.~\ref{eqn:Loss}, the mixed phase is favored when 
\begin{equation}
\sigma \le \frac{n_Q^2}{6\sqrt{3\pi~f_d(x)}~e^2~{\rm d}~  \chi_Q ^{3/2}}\,.
\end{equation}
In the bag model, this condition may be written in terms of $m_s$ and $\mu$ as follows
\begin{equation}
\sigma \lsim 36 ~\left( \frac{m_s}{150 ~{\rm MeV}}\right)^3~\frac{m_s}{\mu}~ {\rm MeV/fm}^2\,. 
\label{eqn:sigma_bag}
\end{equation}
The surface tension between quark matter and vacuum is poorly known. Using the bag model, Berger and Jaffe~\cite{Berger:1986ps} estimate the surface energy of strangelets $\sigma \simeq 8 $ MeV/fm$^2$ for $m_s=150$ MeV and $\mu\simeq 300$ MeV, while $\sigma \simeq 5 $ MeV/fm$^2$ for $m_s=200$ MeV (numerical values for the surface tension are extracted from surface energies quoted in Ref.~\cite{Madsen:1993ka}). The condition     
in Eq.~\ref{eqn:sigma_bag} implies that a structured mixed phase is favored even for $m_s=150$ MeV.  The sensitivity to $m_s$ in Eq.~\ref{eqn:sigma_bag}  and other sources of finite size contributions to the energy which we have neglected here does not allow us to make a definitive claim about the stability of the mixed phase.  Clearly, this warrants further work which should include the curvature energy \cite{Madsen:1993ka}, Debye screening \cite{Norsen:2000wb,Voskresensky:2002hu} and better
estimates of the surface tension. For now, we proceed by assuming that surface and Coulomb costs are small enough to favor the mixed phase.     

\begin{figure}[h]
\includegraphics[width=\columnwidth]{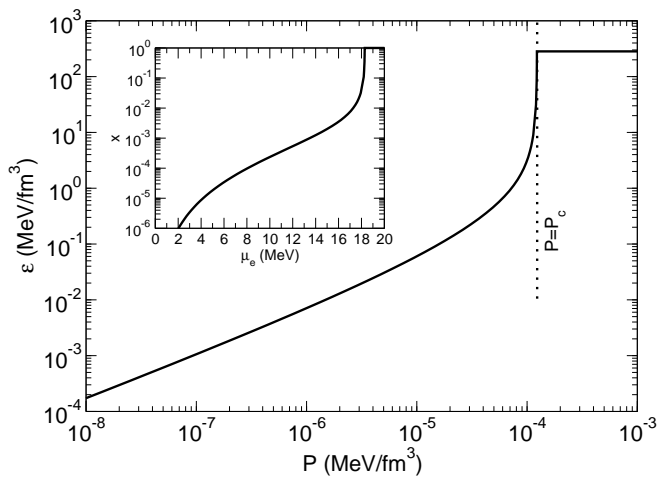}
\caption{The equation of state of the mixed phase. Inset shows the variation of the volume fraction of the quark phase as $\mu_e$ changes across the mixed phase.}
\label{fig:eos}
\end{figure}

We had assumed that $\mu_e^3$ and $\mu_e^4$ terms in the quark pressure were small compared to the electron contribution to facilitate a simplified model independent analysis. We now relax this assumption and work within the Bag model, retaining terms to all orders in $m_s$ and $\mu_e$.  For $B=65$ MeV/fm$^3$ and $m_s=150$ MeV, the quark component of the pressure of homogeneous matter is zero when $\mu = \mu^c \simeq 300$ MeV and $\mu_e=\mu_{e}^{c}\simeq 18$ MeV. The critical pressure below which homogeneous quark matter cannot exist is given by $P_c=(\mu^c_e)^4/12 \pi^2 \simeq 1.2 \times 10^{-4}$ MeV/fm$^3$. The electron chemical potential decreases from $\mu_e=\mu_e^c$ at the critical point to zero at zero pressure.  In the mixed phase the pressure is due to electrons, and is given by 
$P_{\rm mixed}= \tilde{\mu}_e^4/12\pi^2$ while the energy density is due to nuggets, and is given by $\epsilon_{\rm mixed}=x~\epsilon_0$. The variation of pressure and energy density across the mixed phase  are shown in  Fig.~\ref{fig:eos}.
%==============================================
%===========================================         
For our choice of bag model parameters, we find $\epsilon_0\simeq 283$ MeV/fm$^3$. The inset in Fig.~\ref{fig:eos} shows how the volume fraction of the quark phase changes with $\mu_e$. Near the critical point, $x$ and consequently the EoS through $\epsilon_{\rm mixed}= x ~\epsilon_0$, varies rapidly. Except for the region very close to $P_c$, most of the mixed phase is characterized by small $x$. Here, sparsely distributed spherical nuggets are preferred.

\begin{figure}[h]
\includegraphics[width=\columnwidth]{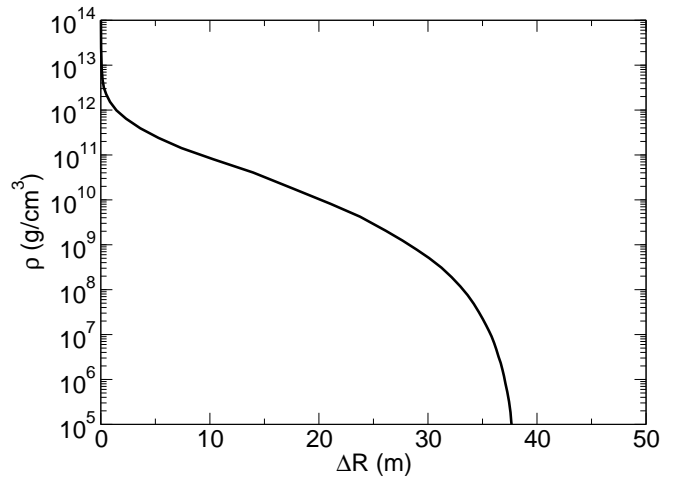}
\caption{Density profile of the crust for a strange star with mass $M=14~M_\odot$ and radius $R=10$ km.}
\label{fig:surface}
\end{figure}

To estimate the radial extent  $\Delta R$ of the mixed phase crust, consider a strange star with mass $M$ and radius $R$. For simplicity, ignoring special and general relativistic corrections, and using Newtonian approximations to hydrostatics
\begin{eqnarray}
GM\int^{R+\Delta R}_{R}\frac{dr}{r^2}&=& \int^{P=0}_{P_c}~\frac{dP}{\epsilon_{\rm mixed} } \,, \\
{\rm when} ~\Delta R \ll R \,,\quad 
\Delta R &=&\frac{R^2}{G M} ~\int^{P_c}_{0}~\frac{dP}{\epsilon_{\rm mixed} } \,,\end{eqnarray}
where $P_c$ is the critical pressure at which the transition to the mixed phase occurs. Since the energy density $\epsilon_{\rm mixed} = x~\epsilon_{\rm 0}$ and the pressure $P=\mu_e^4/12\pi^2$, we may use Eq.~\ref{eqn:volfrac} to obtain
\begin{eqnarray}   
\Delta R &=& \frac{R^2}{G M}~\frac{n_Q}{\epsilon_{\rm 0}}~  \int^{{\mu}^c_e}_{0} d\mu_e~\left(1-\frac{\chi_Q \mu_e}{n_Q}\right) \,,\\
&=& \frac{R}{R_s}~\frac{n_Q^2}{\chi_Q \epsilon_{\rm 0}}~R\,,
\label{eqn:analyticdr}
\end{eqnarray}
where $R_s=2 G M\simeq 3 (M/M_\odot)$ km is the Schwarzschild radius of
the star. For $m_s=150$ MeV and $\mu^{c}\simeq 300
$ MeV we find that $n_Q\simeq0.045$ fm$^{-3}$, $\chi_Q\simeq92$
MeV/fm, and $\epsilon_{\rm 0}\simeq 283$ MeV/fm$^3$. Substituting
these value in Eq.~\ref{eqn:analyticdr} we find that $\Delta R\simeq 100$ meters for a star with mass $M=1.4 M_{\odot}$.  The Newtonian estimate for $\Delta R$ which was obtained using the simplified EoS provides useful insight. To obtain a more accurate value for $\Delta R$, we use the bag model EoS shown in Fig.~\ref{fig:eos} and numerically solve general relativistic equations for hydrostatic equilibrium. The resulting density profile of the crust is shown in Fig.~\ref{fig:surface}. Here we find that $\Delta R \simeq 40$ m.  
%==============================================
%==========================================

We expect the presence of a crust with nuggets to have several phenomenological consequences. \\
 (i) {\it Photon radiation}: In recent years there have been several proposals to look for 
interesting spectral features in the photon radiation from bare strange stars \cite{Page:2002bj,Jaikumar:2004rp}. In the conventional picture, the high electron chemical potential at the surface of the star leads to the formation of an electro-sphere with a high plasmon frequency at its inner edge.  This has been shown to result in unique signatures in the photon radiation especially in the gamma ray region. In our scenario the presence of the crust obviates the need for the electro-sphere, or large photon luminosities thereof.  \\
(ii) {\it Moment of Inertia}: We find that  the moment of inertia of the solid crust $I_{\rm crust}  \sim I_0~M_{\rm crust} / M_{\rm star}$ where $I_0$ is the moment of inertia of the star and 
$ M_{\rm crust} /M_{\rm star}  \sim 10^{-3}-10^{-2}$ is the fractional mass in the crust. This relatively large crustal moment of inertia would imply that a "star-quake" in the nugget crust, which changes its moment of inertia by one part in $10^{4}-10^{6}$, could account for the observed spin-up in the Vela and Crab pulsars. \\
(iii){\it Thermal Conductivity}:  The small mean free path for electrons scattering off nuggets implies that the thermal conductivity in the crust is much smaller than in the core. Much work has been done on calculating the thermal conductivity in the nuclear crust \cite{Baiko:1998xk}. Using the approximate expressions in~\cite{Gnedin:2000me}, we find that the quark crust has a thermal conductivity only slightly less than the nuclear crust. This is relevant to the thermal evolution of such stars, since the crust will act as an insulator effectively keeping the surface temperature low even while the core is hot. Scattering off nuggets is also likely to impact  neutrino transport during the early evolution of the strange star subsequent to its birth in a supernova event\cite{Reddy:1999ad}.   

To reiterate our main findings, a homogeneous and locally charge neutral phase of quarks and electrons could become unstable to phase separation at small pressure. In strange stars, this will lead to the formation of a heterogeneous solid crust where strange nuggets are embedded in a degenerate electron gas. Such a crust shares several similarities to the conventional nuclear crust on normal neutron stars, and can drastically alter the thermal and transport properties of the surface. An interesting exception is stable color-flavor-locked (CFL) quark matter \cite{Alford:1998mk}. In this phase Cooper pairing between up, down and strange quarks leading to color superconductivity ensures that their numbers are equal~\cite{Rajagopal:2000ff,Steiner:2002gx}. The energy gain due to pairing compensates for the larger strange quark mass. The CFL phase is neutral without electrons and stars made entirely of CFL matter would not have a crust of the type proposed here. 

Ultimately, the question of whether or not strange stars have strange crusts depends on the value of the surface tension between strange quark matter and the vacuum. To establish that large strange nuggets are indeed unstable with respect to fission at low pressure we also need to properly account for the role of Debye screening and curvature energy. These finite size contributions to the energy and their model dependence are currently being investigated. They will be reported elsewhere. If these are small enough, then almost all strange stars should have a crust and strange nuggets at zero pressure should have a finite stable size.
%==============================================
%\begin{figure}[ht]
%\includegraphics[width=\columnwidth]{}
%\caption{}
%\label{nearpc}
%\end{figure}
%===========================================

{\it Acknowledgments:} 
We thank Mark Alford, Krishna Rajagopal and Fridolin Weber for useful discussions.  The work of S. R. and A. S. was supported by the Dept. of Energy under contract W-7405-ENG-36.  P. J. is supported by the Department of Energy, Office of Nuclear Physics, contract nos. W-31-109-ENG-38.

%=========== Bibliography ==================

%\bibliographystyle{h-physrev4}
%\bibliography{notsostrangestars} 
%===========================================
\end{document}